\documentclass[twocolumn,showpacs,aps,prl,amsmath,amssymb,
superscriptaddress,letterpaper]{revtex4}

\usepackage{graphicx}
\usepackage{dcolumn}
\usepackage{bm}

\newcommand{\tabref}[1]{Table~\ref{#1}}  
\def\etal{{\em et al.}}

\def\journal#1#2#3#4{{\em #1}~{\bf #2}, #3 (#4)}
\def\PL#1#2#3{\journal{Phys.\ Lett.}{#1}{#2}{#3}}
\def\NP#1#2#3{\journal{Nucl.\ Phys.}{#1}{#2}{#3}}
\def\PR#1#2#3{\journal{Phys.\ Rev.}{#1}{#2}{#3}}
\def\PRL#1#2#3{\journal{Phys.\ Rev. Lett.}{#1}{#2}{#3}}

\def\NIM#1#2#3{\journal{Nucl.\ Instrum.\ Methods}{#1}{#2}{#3}}

\newcommand{\Moller}{M{\o}ller}
\newcommand{\qwe}{Q_W^e}
\newcommand{\stwmsbar}
{\sin^2\theta_{\mathrm{W}}(M_{\rm{Z}})_{\overline{\rm{MS}}}}

\begin{document}


\title{Observation of Parity Nonconservation in \Moller\ Scattering}

\affiliation{University of California, Berkeley, California 94720}

\affiliation{California Institute of Technology, Pasadena,
California 91125}
\affiliation{University of Massachusetts, Amherst, Massachusetts 01003}

\affiliation{Princeton University, Princeton, New Jersey 08544}

\affiliation{CEA Saclay, DAPNIA/SPhN, F-91191 Gif-sur-Yvette, France}

\affiliation{Smith College, Northampton, Massachusetts 01063}

\affiliation{Stanford Linear Accelerator Center, Menlo Park,
California 94025}

\affiliation{Syracuse University, Syracuse, New York  13244}

\affiliation{Thomas Jefferson Laboratory, Newport News, Virginia
23606}

\affiliation{University of Virginia, Charlottsville, Virginia
22903}

\author{P.L.~Anthony}\affiliation{Stanford Linear Accelerator Center, Menlo Park, California 94025}
\author{R.G.~Arnold}\affiliation{University of Massachusetts, Amherst, Massachusetts 01003}
\author{C.~Arroyo}\affiliation{University of Massachusetts, Amherst, Massachusetts 01003}
\author{K.~Baird}\affiliation{Stanford Linear Accelerator Center, Menlo Park, California 94025}
\author{K.~Bega}\affiliation{California Institute of Technology, Pasadena, California 91125}
\author{J.~Biesiada}
\affiliation{University of California, Berkeley, California 94720}
\affiliation{Princeton University, Princeton, New Jersey 08544}
\author{P.E.~Bosted}\affiliation{University of Massachusetts, Amherst, Massachusetts 01003}
\author{M.~Breuer}\affiliation{University of Massachusetts, Amherst, Massachusetts 01003}
\author{R.~Carr}\affiliation{California Institute of Technology, Pasadena, California 91125}
\author{G.D.~Cates}\affiliation{University of Virginia, Charlottsville, Virginia 22903}
\author{J-P.~Chen}\affiliation{Thomas Jefferson Laboratory, Newport News, Virginia 23606}
\author{E.~Chudakov}\affiliation{Thomas Jefferson Laboratory, Newport News, Virginia 23606}
\author{M.~Cooke}\affiliation{University of California, Berkeley, California 94720}
\author{F.J.~Decker}\affiliation{Stanford Linear Accelerator Center, Menlo Park, California 94025}
\author{P.~Decowski}\affiliation{Smith College, Northampton, Massachusetts 01063}
\author{A.~Deur}\affiliation{University of Virginia, Charlottsville, Virginia 22903}
\author{W.~Emam}\affiliation{Syracuse University, Syracuse, New York  13244}
\author{R.~Erickson}\affiliation{Stanford Linear Accelerator Center, Menlo Park, California 94025}
\author{T.~Fieguth}\affiliation{Stanford Linear Accelerator Center, Menlo Park, California 94025}
\author{C.~Field}\affiliation{Stanford Linear Accelerator Center, Menlo Park, California 94025}
\author{J.~Gao}\affiliation{California Institute of Technology, Pasadena, California 91125}
\author{K.~Gustafsson}
\altaffiliation[Now at: ]{Helsinki Institute of Physics, Finland}
\affiliation{California Institute of Technology, Pasadena, California 91125}
\author{R.S.~Hicks}\affiliation{University of Massachusetts, Amherst, Massachusetts 01003}
\author{R.~Holmes}\affiliation{Syracuse University, Syracuse, New York  13244}
\author{E.W.~Hughes}\affiliation{California Institute of Technology, Pasadena, California 91125}
\author{T.B.~Humensky}\affiliation{Princeton University, Princeton, New Jersey 08544}
\author{G.M.~Jones}\affiliation{California Institute of Technology, Pasadena, California 91125}
\author{L.J.~Kaufman}\affiliation{University of Massachusetts, Amherst, Massachusetts 01003}
\author{Yu.G.~Kolomensky}\affiliation{University of California, Berkeley, California 94720}
\author{K.S.~Kumar}
\affiliation{University of Massachusetts, Amherst, Massachusetts 01003}
\author{D.~Lhuillier}\affiliation{CEA Saclay, DAPNIA/SPhN, F-91191 Gif-sur-Yvette, France}
\author{R.~Lombard-Nelsen}\affiliation{CEA Saclay, DAPNIA/SPhN, F-91191 Gif-sur-Yvette, France}
\author{P.~Mastromarino}\affiliation{California Institute of Technology, Pasadena, California 91125}
\author{B.~Mayer}\affiliation{CEA Saclay, DAPNIA/SPhN, F-91191 Gif-sur-Yvette, France}
\author{R.D.~McKeown}\affiliation{California Institute of Technology, Pasadena, California 91125}
\author{R.~Michaels}\affiliation{Thomas Jefferson Laboratory, Newport News, Virginia 23606}
\author{M.~Olson}\affiliation{Stanford Linear Accelerator Center, Menlo Park, California 94025}
\author{K.D.~Paschke}\affiliation{University of Massachusetts, Amherst, Massachusetts 01003}
\author{G.A.~Peterson}\affiliation{University of Massachusetts, Amherst, Massachusetts 01003}
\author{R.~Pitthan}\affiliation{Stanford Linear Accelerator Center, Menlo Park, California 94025}
\author{K.~Pope}
\thanks{Deceased}
\affiliation{Smith College, Northampton, Massachusetts 01063}
\author{D.~Relyea}\affiliation{Princeton University, Princeton, New Jersey 08544}
\affiliation{Stanford Linear Accelerator Center, Menlo Park, California 94025}
\author{S.E.~Rock}\affiliation{University of Massachusetts, Amherst, Massachusetts 01003}
\author{O.~Saxton}\affiliation{Stanford Linear Accelerator Center, Menlo Park, California 94025}
\author{G.~Shapiro}
\thanks{Deceased}
\affiliation{University of California, Berkeley, California 94720}
\author{J.~Singh}\affiliation{University of Virginia, Charlottsville, Virginia 22903}
\author{P.A.~Souder}\affiliation{Syracuse University, Syracuse, New York  13244}
\author{Z.M.~Szalata}\affiliation{Stanford Linear Accelerator Center, Menlo Park, California 94025}
\author{W.A.~Tobias}\affiliation{University of Virginia, Charlottsville, Virginia 22903}
\author{B.T.~Tonguc}\affiliation{Syracuse University, Syracuse, New York  13244}
\author{J.~Turner}\affiliation{Stanford Linear Accelerator Center, Menlo Park, California 94025}
\author{B.~Tweedie}\affiliation{University of California, Berkeley, California 94720}
\author{A.~Vacheret}\affiliation{CEA Saclay, DAPNIA/SPhN, F-91191 Gif-sur-Yvette, France}
\author{D.~Walz}\affiliation{Stanford Linear Accelerator Center, Menlo Park, California 94025}
\author{T.~Weber}\affiliation{Stanford Linear Accelerator Center, Menlo Park, California 94025}
\author{J.~Weisend}\affiliation{Stanford Linear Accelerator Center, Menlo Park, California 94025}
\author{D.~Whittum}\affiliation{Stanford Linear Accelerator Center, Menlo Park, California 94025}
\author{M.~Woods}\affiliation{Stanford Linear Accelerator Center, Menlo Park, California 94025}
\author{I.~Younus}\affiliation{Syracuse University, Syracuse, New York  13244}

\collaboration{SLAC E158 Collaboration}\noaffiliation

\date{December 10, 2003}

\begin{abstract}
We report a measurement of the parity-violating asymmetry
in fixed target electron-electron (\Moller) scattering: 
$A_{PV} = (-175\pm 30\ ({\rm stat.})\pm 20\ ({\rm syst.}))
\times 10^{-9}$. This
first direct observation of parity nonconservation in \Moller\ scattering
leads to a measurement of the electron's weak charge 
at low energy $\qwe=-0.053\pm 0.011$. This 
is consistent with the Standard Model expectation 
at the current level of precision:
$\stwmsbar = 0.2293\pm 0.0024\ \mathrm{(stat.)}\pm 0.0016\ \mathrm{(syst.)
\pm 0.0006\ \mathrm{(theory)}}$.
\end{abstract}

\pacs{11.30.Er, 12.15.Lk, 12.15.Mm, 13.66.Lm, 13.88.+e, 14.60.Cd}
\maketitle

Precision measurements of weak neutral current (WNC) processes
mediated by Z$^0$ exchange
stringently test the Standard Model of electroweak interactions.
While most WNC measurements have been performed at 
high energy colliders, the comprehensive search for
new physics at TeV energies also requires precision measurements
at low momentum transfer ($Q^2\ll M_Z^2$). 

One class of such measurements involves the scattering of longitudinally
polarized electrons from unpolarized targets, 
allowing the determination of a parity-violating
asymmetry $A_{PV}\equiv{(\sigma_R-\sigma_L)}/{(\sigma_R+\sigma_L)}$,
where $\sigma_{R(L)}$ is the cross section for incident right(left)-handed 
electrons.
$A_{PV}$ arises from the interference of the weak and electromagnetic
amplitudes \cite{zeld} and is sensitive to WNC coupling constants and thus 
the weak mixing angle $\theta_{\rm W}$.

The first observation of $A_{PV}$ was made 
at the Stanford Linear Accelerator Center (SLAC) using a deuteron 
target \cite{e122}.
That experiment established the basic experimental technique to determine
small asymmetries which typically range from 0.1 to 100 parts per million 
(ppm). Subsequent measurements yielded improved 
precision and accuracy \cite{expt,kkreview}. However, 
theoretical uncertainties related to the use of hadronic targets 
restricted the interpretation of the experimental
results at the quantum loop level.

In this paper we report the first observation of $A_{PV}$ in electron-electron
(\Moller) scattering. This purely leptonic reaction has a large cross section
and has little theoretical uncertainty.
The development of the 50 GeV electron beam in SLAC End Station A (ESA) made 
possible a measurement of $A_{PV}$ \cite{modphy} with a precision that tests
electroweak radiative corrections
and probes physics beyond the Standard Model at the TeV scale.


At 50 GeV and a center-of-mass scattering angle of $90^\circ$,
$A_{PV}$ in \Moller\ scattering is predicted to be $\simeq 320$ 
parts per billion (ppb) \cite{darman} at tree level. Electroweak 
radiative corrections 
\cite{marciano1,otherad}
and the experimental acceptance reduce the measured asymmetry 
by more than 50\%.
The principal components of the experimental apparatus, designed to measure
$A_{PV}$ to better than 10\%, were the
polarized electron
beam, a liquid hydrogen target, a spectrometer/collimator system, and 
detectors. 
\Moller-scattered electrons were directed into a calorimeter by a magnetic
spectrometer.
The asymmetry was measured by extracting the fractional difference in the 
integrated calorimeter response for incident right- and left-handed 
beam pulses.

\begin{figure*}
\includegraphics[width=6.5in]{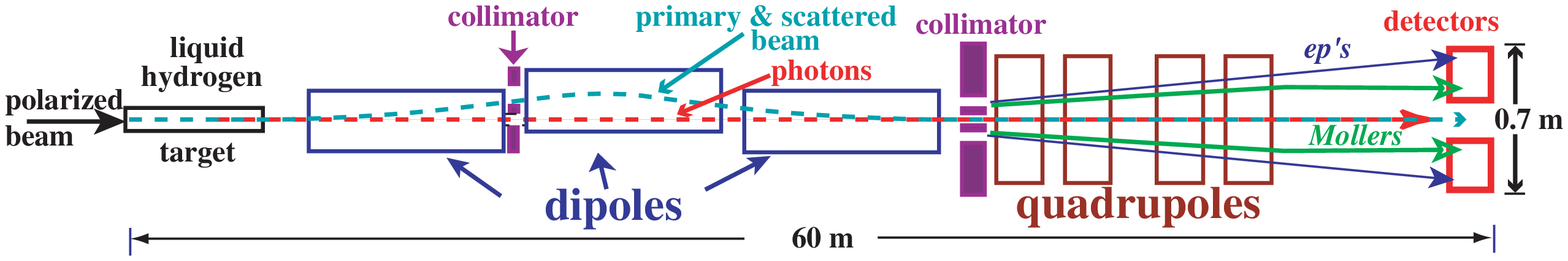}
\caption{\label{overview} Schematic plan view of the experimental 
configuration in SLAC End Station A.
}
\end{figure*}
The longitudinally polarized electron beam, with up to $5.5\times 10^{11}$
electrons in $\approx 270$~ns pulses at 120~Hz, was 
produced by optical pumping of a strained GaAs photocathode \cite{takashi} 
by circularly polarized laser light \cite{brian}. The sign 
of the laser circular polarization state determined the electron beam helicity.
The helicity sequence of the pulse train was made up of 
quadruplets consisting of two consecutive 
pulses with pseudo-randomly chosen helicities, 
followed by their complements, yielding
two independent right-left ``pulse pairs" every 33~ms. 

Careful optimization of optical components \cite{brian} minimized intrinsic 
intensity and position differences between right- and left-helicity beams
that resulted from 
imperfections in the laser light and the photocathode response. 
Additionally, helicity-dependent 
corrections were applied to the laser beam in a feedback loop using
periodic average measurements of beam asymmetries.
The beam intensity and position were measured at the 
upstream and downstream ends of the accelerator with typical accuracies 
per pulse pair of 50~ppm in intensity, 50~ppm in energy 
and 2~$\mu$m \cite{yury} in position. 
Cumulative beam asymmetries at the target were reduced to $<500$~ppb 
in intensity, $<10$~ppb in energy, and $<50$~nm in position.


A schematic diagram of the apparatus in ESA is shown in 
Fig.~\ref{overview}. 
The 0.5~MW electron beam first passed through a 1.57~m long
cylindrical cell filled with liquid hydrogen \cite{target}. 
The cell was part of a target loop
consisting of a motor, impeller, and heat exchanger, which circulated the
liquid hydrogen at $\approx 5$~m/s. 
Aluminum meshes in the fluid path surrounding the electron beam enhanced 
turbulence
and mixing. These features allowed the absorption of $\approx 500$~W 
deposited by
the beam while keeping density fluctuations below 40~ppm per pulse
pair. 

Scattered particles with laboratory scattering angle between 4.4 and 7.5 mr
over the
full range of the azimuth
were selected by the magnetic spectrometer \cite{davidthesis}
while the primary beam and 
forward angle photons passed unimpeded to the beam dump.
Sixty meters downstream of the target, the charged particle flux was 
approximately azimuthally symmetric about the beam axis. 
\Moller\ electrons in the range 13-24~GeV formed a ring
spatially separated from electrons 
scattered from target protons ($ep$ scattering). 
This \Moller\ ring contained $\approx 2\times 10^{7}$ electrons per pulse.

\begin{figure}
\includegraphics[width=3.3in]{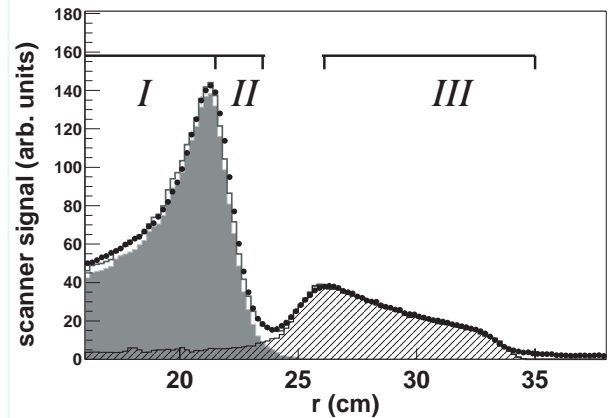}
\caption{\label{fig_profile} The charged particle radial profile
at the calorimeter. The points are the data scan, and the 
open histogram is the Monte Carlo simulation.  
\Moller\ (shaded) and $ep$ (hatched) contributions are also shown.
Region {\it I} and {\it III}
PMTs were used to measure \Moller\ and $ep$ asymmetries, respectively.
}
\end{figure}

A scanning detector system provided a 
complete radial and azimuthal map of the charged particle flux.
Figure~\ref{fig_profile} compares the measured radial profile at one
azimuthal angle with Monte Carlo simulations of \Moller\ and $ep$ scattering.
The $ep$ flux within the \Moller\ ring was the dominant background, 
estimated to be $\simeq 8$\%.

For the asymmetry measurement, the charged particle 
flux was intercepted by the primary calorimeter, 
a 25~cm long cylindrical structure with a 15(35)~cm inner(outer) radius. 
It was assembled by layering planes of flexible fused-silica fibers between
elliptical copper plates so as to withstand a 100~Mrad radiation dose.
The fibers directed Cherenkov light to air light-guides,
each of which terminated into a shielded photomultiplier tube (PMT). 
The regions {\it I}, {\it II} and {\it III}
in Fig.~\ref{fig_profile} were covered by 30, 20 and 10 PMTs, respectively,
providing radial and azimuthal segmentation. 

The small contribution of neutral particles, such as photons and
neutrons, to the calorimeter response 
was measured in calibration runs. 
The asymmetry from pions was measured by using ten quartz bars arranged in 
azimuthal symmetry behind the \Moller\ detector and lead shielding.
Eight ionization chambers arranged in $45^\circ$
azimuthal sections intercepted charged particles with  
$\theta_{\mathrm {lab}}\approx 1$~mr.  
This ``luminosity" detector monitored target 
density fluctuations and provided a check of the null asymmetry expected
at such small scattering angles.


The data sample, constituting a total flux of just over
$10^{20}$ electrons on target,
was collected at beam energies of 45.0 and 48.3 GeV. Due to
$g-2$ precession as the beam traversed a $24.5^\circ$ bend
after acceleration,
the two beam energies corresponded
to opposite orientations of longitudinal beam polarization in ESA. 
Roughly equal statistics were thus accumulated with 
opposite signs for the measured asymmetry,
which suppressed many systematic effects.
In addition, the state of a half-wave plate in the laser line was
toggled every other day, passively reversing the sign of the electron
beam polarization. This guarded against helicity-correlated
electronics crosstalk.


For each beam pulse at 120 Hz, 
a distributed data acquisition (DAQ) system was triggered
to collect data from the polarized source electronics and the digitized
integrated response of the detectors and beam monitors with negligibly small
electronic dead time.   
Alternate DAQ triggers fell into two 60~Hz fixed-phase ``time slots". Within
these time slots, right-left pulse pairs were formed for independent asymmetry
analyses.

Loose requirements were imposed on beam quality and beam monitor linearity.
However, no helicity-dependent cuts were applied, other than the demand that
the beam intensity asymmetry measured by two independent monitors agreed to 
within $10^{-3}$ for each pulse-pair.
In total,
$8.6\times 10^7$ pulse pairs satisfied all selection criteria.
The right-left asymmetry in the integrated detector response for each pulse
pair was computed and 
then corrected for fluctuations in the beam trajectory.

To first
order, six correlated beam parameters described the
trajectory of a beam pulse: intensity, energy, and horizontal and vertical
position and angle. Each beam parameter was measured by two
independent monitors, such that device resolution and systematic
effects could be studied. 

Two methods were used to calibrate
the detector sensitivity to each beam parameter and thus remove
beam-induced random and systematic effects from the raw asymmetry. 
One method used a calibration subset of the pulses,
where each beam parameter was modulated periodically
around its average
value by an amount large compared to nominal beam fluctuations with a 
$\sim$4\% duty cycle. 
The other method applied an unbinned least squares linear regression 
to the pulses used for physics analysis.
They yielded statistically consistent results to within 12 ppb.
Final results were obtained with the latter, statistically more powerful
technique. 

The integrated responses of region {\it I} PMTs (see 
Fig.~\ref{fig_profile})
were averaged to form the raw asymmetry $A_{\mathrm{raw}}$.
Near-perfect azimuthal symmetry reduced the sensitivity to beam fluctuations
and right-left asymmetries. The $A_\mathrm{raw}$ pulse-pair distribution 
had an RMS of $\approx 200$~ppm. The cumulative beam asymmetry correction 
was $-41\pm 3$~ppb.
A correction due to an azimuthal modulation of 
$A_{\mathrm{raw}}$ \cite{transverse}
from a small non-zero transverse beam polarization component 
was found to be $-8\pm 3$~ppb.

Additional bias to $A_{\mathrm{raw}}$ may arise from
asymmetries in unmeasured beam quantities, such as higher order
moments of beam distributions. Region {\it II} PMT channels
were significantly more sensitive to such fluctuations and helped to limit 
possible bias in $A_{\mathrm{raw}}$ to 10~ppb. 
Likewise, the luminosity detector was very sensitive 
to higher order effects, where the cumulative raw right-left
asymmetry was $[-16\pm 15\rm{(stat.)}]$~ppb. This
is consistent with the theoretical expectation, providing additional 
confirmation
that higher order effects are under control.
A separate study limited the 
bias due to beam spot-size fluctuations on $A_\mathrm{raw}$ to 1~ppb, using
data from a retractable wire array that was inserted into the beam during
some of the data collection. 

The physics asymmetry $A_{\rm phys}$ was formed from $A_{\rm raw}$ by
correcting for background
contributions, detector linearity and beam polarization:
\[
A_{\rm phys} = \frac{1}{P_b \epsilon} 
\frac{A_{\rm raw} - \sum_i \Delta A_i}{1-\sum_i f_i} \, .
\]
$\Delta A_i$ and dilutions $f_i$ for various background sources are
listed in \tabref{tab:systematics}. 
\begin{table}
\caption{Corrections $\Delta A_i$ and dilutions $f_i$ 
to $A_{\rm raw}$ and associated systematic uncertainties.}
\label{tab:systematics}
\begin{center}
\begin{tabular}{|l|r|c|}
\hline
Source & $\Delta A$ (ppb) & $f$ \\
\hline
Beam (first order)  & $-41\pm 3$& \\
Beam (higher order) & $0\pm 10$& \\
Transverse polarization & $-8\pm 3$ & \\
$e^-+p\rightarrow e^-+p(+\gamma)$ & $-8\pm 2$ & $0.064\pm 0.007$ \\
$e^-(\gamma)+p\rightarrow e^-+X$ & $-26\pm 6$ & $0.011\pm 0.003$ \\
High energy photons & $3\pm 3$ & $0.004\pm 0.002$\\
Synchrotron photons & $0\pm 5$ & $0.002\pm 0.001$\\
Neutrons            & $-5\pm 3$ & $0.003\pm 0.001$\\
Pions               & $1\pm 1$  & $0.001\pm 0.001$\\
\hline
\end{tabular}
\end{center} 
\end{table}
The largest correction of $-26\pm 6$~ppb 
was due to electrons from inelastic electron- and photon-proton interactions.
The measured asymmetry in region {\it III} PMTs was used as input, along
with reasonable assumptions for the kinematic extrapolation to region {\it I}.

The electron beam polarization was
$P_b=0.85\pm 0.05$, measured every other day by a dedicated
polarimeter using \Moller\ scattering of beam electrons
off a magnetized foil placed just upstream of the hydrogen target. 
The linearity of the calorimeter response
was determined to be $\epsilon=0.99\pm 0.01$.



\begin{figure}
\includegraphics[width=3.3in]{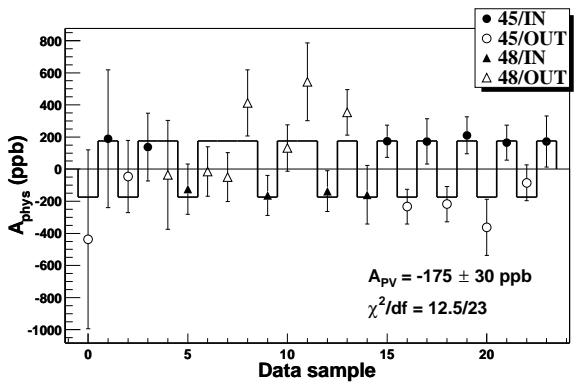}
\caption{\label{fig:slug} $A_{\rm phys}$ for each of 24
data samples. Data collected with half-wave plate inserted(removed) 
at a beam energy of 45(48) GeV are shown
as solid(open) circles(triangles). The solid line represents the grand
average, with the expected modulation of the asymmetry sign 
for each beam energy and half-wave plate state. 
Only statistical uncertainties are shown. 
}
\end{figure}

Figure~\ref{fig:slug} shows $A_{\rm phys}$ for all data, 
divided into 24 sequential samples. Each $A_{\rm phys}$ measurement has 
sign reversals depending on the
beam energy and the state of the half-wave plate.
$A_{PV}$ is obtained by correcting each result by the appropriate sign. 
The combined result is
\[A_{PV} = -175\pm 30\ ({\rm stat.})\pm 20\ ({\rm syst.})~\mathrm{ppb},\] 
establishing parity nonconservation in \Moller\ scattering
at the $5\sigma$ level. 
$A_{PV}$ is proportional to the product of the electron's vector
and axial-vector neutral current couplings, parameterized as 
the weak charge $\qwe$:
%
\[
A_{PV} = \frac{G_F Q^2}{\sqrt{2}\pi\alpha}\frac{1-y}{1+y^4+(1-y)^4}
{\mathcal F}_{\rm b}\qwe \equiv {\mathcal A}(Q^2,y)\qwe,
\]
%
where $G_F$ and $\alpha$ are the Fermi and fine structure constants,
respectively \cite{PDG2002}, and 
${\mathcal F}_{\mathrm b}=1.01\pm 0.01$ accounted for
kinematically weighted hard initial and final state radiation 
effects \cite{Zykunov}.
The effective analyzing power 
${\mathcal A} = 3.28\pm 0.06$~ppm 
was determined from a Monte Carlo simulation 
that accounted for
energy losses in the target and systematic uncertainties in the
spectrometer setup. The average values of the kinematic variables were 
$Q^2 = 0.026~(\mathrm{GeV/c})^2$ and $y\equiv Q^2/s\simeq 0.6$, where $s$
is the square of the center-of-mass energy.

We find 
$\qwe 
= -0.053\pm 0.009\ ({\rm stat.})\pm 0.006\ ({\rm syst.})$,
consistent with the Standard Model expectation \cite{marciano1,PDG2002} of
$-0.046\pm 0.003$. As an example of the sensitivity of the measurement, the 
result can be used to limit the scale $\Lambda_\mathrm{LL}$
of a new left-handed contact interaction characterized by a term in the 
Lagrangian \cite{eichten, mikelambda} 
${\mathcal L}=(4\pi/2\Lambda_\mathrm {LL}^2)(\bar{e}_L\gamma_\mu e_L)$.
At 95\%\ C.L. a tree-level calculation yields
$\Lambda_\mathrm {LL}^+\ge 7.2$ TeV and $\Lambda_\mathrm {LL}^-\ge 5.1$ TeV, 
for potential positive and negative deviations, respectively. 

In the context of the Standard Model, we find
\begin{eqnarray*}
\stwmsbar = 0.2293 & \pm & 0.0024\ ({\rm stat.})\\
          \pm\ 0.0016\ ({\rm syst.}) & \pm &  0.0006\ ({\rm theory}).
\end{eqnarray*}
The last error comes from the $Q^2$ evolution to the Z-pole. 


The reported $A_{PV}$ result is the most precise measurement of any asymmetry
in electron scattering. 
The consistency of the result
with the theoretical prediction provides significant new limits on TeV scale
physics, comparable in sensitivity and complementary to 
other WNC measurements at 
low $Q^2$ \cite{csnutev}. 
Data from the first of three E158 run periods were
used; the measurement accuracy is expected to improve by more than a factor of 
two when analysis of all the data is complete.
The 
experimental techniques described in this Letter
demonstrate the feasibility of measuring asymmetries with
accuracies better than 10~ppb, applicable to new experiments
under development \cite{kkreview}.



\begin{acknowledgments}

We thank the SLAC staff
for their efforts in helping develop and operate the E158 apparatus, and
especially A. Brachmann and T. Maruyama for their essential work on the
polarized electron source. 
This work was supported by Department of Energy contract DE-AC03-76SF00515,
and by the Division of Nuclear Physics at the Department of Energy and 
the Nuclear Physics Division of the National Science Foundation in the United
States and the 
Commissariat \`{a} l'\'{E}nergie Atomique and the Centre National
de la Recherche Scientifique in France.

\end{acknowledgments}

\end{document}